\documentclass[conference]{IEEEtran}
\IEEEoverridecommandlockouts
\usepackage{cite}
\usepackage{amsmath,amssymb,amsfonts}
\usepackage{graphicx}
\usepackage{multirow}
\usepackage{textcomp}
\usepackage{fancyhdr} 
\usepackage{cleveref}
\usepackage{booktabs}
\usepackage{hyphenat}
\usepackage[english]{babel}
\usepackage{float}
\useshorthands{~}
\usepackage[top=0.75in,bottom=1.29in,left=0.5in,right=0.5in]{geometry}
\title{Calculation of Femur Caput Collum Diaphyseal angle for X-Rays images using 
Semantic Segmentation}

\author{
\IEEEauthorblockN{Muhammad Abdullah, Anne Querfurth, Deepak Bhatia, Mahdi Mantash}
\IEEEauthorblockA{University of Erlangen-Nürnberg \\
muhammad.a.abdullah@fau.de, anne.querfurth@fau.de, deepak.bhatia@fau.de, mahdi.man.mantash@fau.de}

}

\begin{document}
\maketitle
%%%%%%%%%%%%%%%%%%%%%%%%%

\begin{abstract}
This paper investigates the use of deep learning approaches to estimate the femur caput-collum-diaphyseal (CCD) angle from X-ray images. The CCD angle is an important measurement in the diagnosis of hip problems, and correct prediction can help in the planning of surgical procedures. Manual measurement of this angle, on the other hand, can be time intensive and vulnerable to inter-observer variability. In this paper, we present a deep-learning algorithm that can reliably estimate the femur CCD angle from X-ray images. To train and test the performance of our model, we employed an X-ray image dataset with associated femur CCD angle measurements. Furthermore we built a prototype to display the resulting predictions and to allow the user to interact with the predictions. As this is happening in a sterile setting during a surgery, we expanded our interface to the possibility of being used only by voice commands.
 Our results show that our deep learning model predicts the femur CCD angle on X-ray images with great accuracy, with a mean absolute error of 4.3 degrees on left femur and 4.9 degrees on right femur on test dataset. Our results suggest that deep learning has the potential to give a more efficient and accurate technique for predicting the femur CCD angle, which might have substantial therapeutic implications for the diagnosis and management of hip problems.

\end{abstract}

\begin{IEEEkeywords}
		biomedical imaging, fully convolutional neural network, femur imaging, heatmap regression, ransac, speech-to-text
		\end{IEEEkeywords}

\section{Introduction} \label{introduction}
Hip fractures are a common injury among the elderly, with high morbidity and death rates. Correct hip fracture diagnosis and treatment are crucial for achieving positive clinical results. The caput-collum-diaphyseal (CCD) angle, which measures the angle formed between the femoral head and neck and the femoral shaft, is one of the major factors utilized in the diagnosis and therapy of hip fractures. The accurate measurement of the CCD angle is critical for identifying the best treatment strategy for hip fractures, including implant selection and surgical technique. \cite{hip_fracture}
Manual CCD angle measurement by radiologists can be time-consuming and vulnerable to inter-observer variability.\\
Deep learning-based algorithms have shown promising outcomes in medical image analysis applications in recent years. Convolutional neural networks (CNNs) are used in these approaches to extract features. The effectiveness of these approaches, however, is largely dependent on the quality and amount of the training dataset. We present a deep learning-based technique for estimating the CCD angle on X-ray images in this research. Our solution uses a U-Net architecture \cite{unet} to learn information from X-ray images and predict the CCD angle. The suggested method is evaluated using a set of X-ray images of patients that have been labelled with the associated CCD angles. With an MAE of 4.3 degrees on left femur and 4.9 degrees on right femur, our experimental findings suggest that the proposed technique performs quite well in terms of accuracy. 
\newline
The remainder of this paper is organized as follows. Section II discusses the theoretical aspects of the research, such as the segmentation model utilized and the tasks performed, as well as the speech recognition model deployed. We describe the proposed method in section III, including the dataset used for training and evaluation, the training procedure, evaluation, and the design and functioning of the user interface of our prototype. Section IV presents the experimental findings. Section V, we conclude and discusses the advantages and limits of the proposed technique. Finally, Section VI suggests future research areas.

\section{Theory} \label{Theory}
\subsection{Segmentation model}
Image segmentation is the process of classifying each pixel to a class. It can be further divided into \textit{instance segmentation} and \textit{semantic segmentation}. Semantic segmentation cannot distinguish between different instances of the same class while instance segmentation is capable of doing it. \newline
Image segmentation has many applications in medical imaging where it is used to identify irregular structures and abnormalities. Recent developments in deep-learning also entail performance enhancement of image segmentation networks. Many 
new deep-learning-based segmentation architectures have been proposed in the recent past. It includes Unet\cite{unet}, MASK-RCNN\cite{he2018mask}, SegNet \cite{badrinarayanan2016segnet}, DeepLabv3+\cite{chen2018encoderdecoder}. After the success of transformers for natural language tasks, they have also been introduced for vision tasks as well where they are labeled as vision transformers. \cite{liu2021swin}\textit{Swin Transformer} is one such example. Our work does not include performance comparison of different segmentation architectures.
\newline
We decided to use UNet as it generalizes well even with the limited amount of training data. Since training data is always sparse in medical imaging tasks, UNet has been widely used in medical imaging for segmentation tasks.
\newline
UNet is a fully convolutional network\cite{long2015fully}. It includes a contracting path for context learning and a symmetric expanding path for precise localisation. Localisation is important in biomedical images. 2D transposed convolution is used for upsampling in expanding paths.

\begin{figure}[ht]
\centering
    \includegraphics[width=\linewidth]{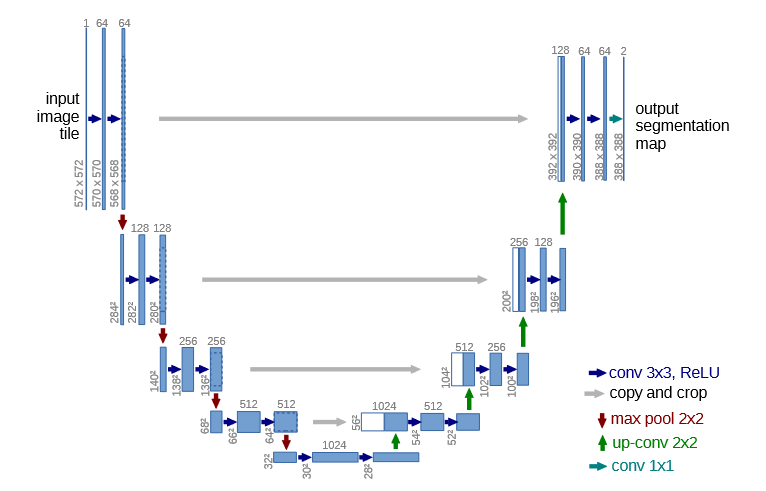}
\caption{U-shaped UNet architecture} 
\label{fig:unet}
\end{figure}
\subsection{Segmentation task}
Our goal is to calculate the CCD angle. CCD angle is measured between the femur neck centerline and the femur shaft centerline. These centerlines can be predicted directly or can be estimated from lateral and medial lines both for the neck and shaft. Moreover, the femur neck centerline always passes through the centre of the femur head. This extra information can also be incorporated for the estimation of the femur neck centerline.
\newline
Lines can be regressed directly or indirectly via segmentation of the Gaussian distributed line heatmap. Heatmap regression has performed better in pose-estimation tasks where the goal is to detect the location of human joints or key points\cite{xiao2018simple}. Heatmap regression learns a probability distribution by taking neighbouring pixels into account. It is more robust and generalizes well.
Our work is also based on regressing lines indirectly via heatmaps.
\newline
Considering the possibility of more than one way to estimate the femur neck centerline and femur shaft centerline, we decided to train a 12 output channel UNet network incorporating 3 lines (medial, centerline, lateral) both for the neck and shaft. We also predict CCD angle both for the left femur and right femur. So, in total, 6 lines for the left femur and 6 lines for the right femur.

\subsection{Motivation for Touchless Interface}
The working environment in which our application is used is the operating room and therefore comes with special challenges. These include limited space, time pressure to keep the patient's anaesthesia as short as possible, and sterility. The interaction with a touch-based system requires the doctor to become non-sterile which leads to the necessity of making oneself sterile again. This costs valuable time and material.\\ 
New technologies can help, as there are already a large number of touch-less input devices that allow interaction without touching. These include eye tracking, gesture recognition and voice recognition.
All of these devices represent a great opportunity to reduce time in the OR and allow a steady workflow. \cite{mewes_touchless_2017}\\
We considered using gesture recognition. However, compared to voice recognition it has many disadvantages. For example, the user would need to have free hands and the correct visibility angle and lighting for the camera to recognise their gestures successfully. Furthermore, using gestures to control an application is not intuitive and requires more time to become proficient and know all of the gestures by heart. Considering all of those error sources, we decided to go with voice recognition instead.\\
Voice recognition as an input method holds further advantages. The additionally required hardware is only a microphone. Nowadays microphones are small, light and provide nevertheless high-quality sound recordings. This makes it possible to be used in a large variety of spatial settings as the surgeon can move around the room while for example having the microphone attached to them.
A relevant factor in employing a new application or workflow in the hospital setting is the intuitivity and simplicity of using and learning how to handle the new system. Voice recognition provides this aspect, as the commands are displayed on the monitor as well as being regular commands that you would give a human to perform the wanted task. The required time to get proficient at using the system is therefore minimal. \\
Overall, voice recognition provides a needed addition to our interface to make it valuable for the OR setting.

\subsection{Voice model}
OpenAI Whisper is a speech recognition deep learning model that combines a convolutional neural network and a transformer-based language model. The CNN component of the model encodes the temporal and spectral properties of the input audio signal, while the transformer-based language model generates transcriptions based on the stored features. \cite{li2020openai} One of OpenAI Whisper's key advantages is its capacity to tolerate loud, noisy and difficult acoustic situations which suit very well to our requirements for a speech recognition model in a busy interventional setting. We used OpenAI Whisper for speech-to-text recognition and implement voice control feature to control our user interface.

\section{Methods} \label{Methods}

\subsection{Data}
The data comprised 201 hip X-ray images of 166 different patients with both left and right femurs. The images were annotated for 18 different labels, 9 for each femur bone at the start of the project by the 8 participants to look for different strategies to  predict the CCD angle. The labels comprised the following: 
Femur Right, 
Femoral Head Right, 
Femoral Head Circle Right, 
Femoral Shaft Medial Right, 
Femoral Shaft Lateral Right, 
Femoral Shaft Centerline Right, 
Femoral Neck Medial Right, 
Femoral Neck Lateral Right, 
Femoral Neck Centerline Right, 
Femur Left, 
Femoral Head Left, 
Femoral Head Circle Left, 
Femoral Shaft Medial Left, 
Femoral Shaft Lateral Left, 
Femoral Shaft Centerline Left, 
Femoral Neck Medial Left, 
Femoral Neck Lateral Left, 
Femoral Neck Centerline Left. Our team found some problems with the neck lines in the annotations so we re-annotated the neck lines.  

\begin{figure}[ht]
\centering
    \includegraphics[width=\linewidth]{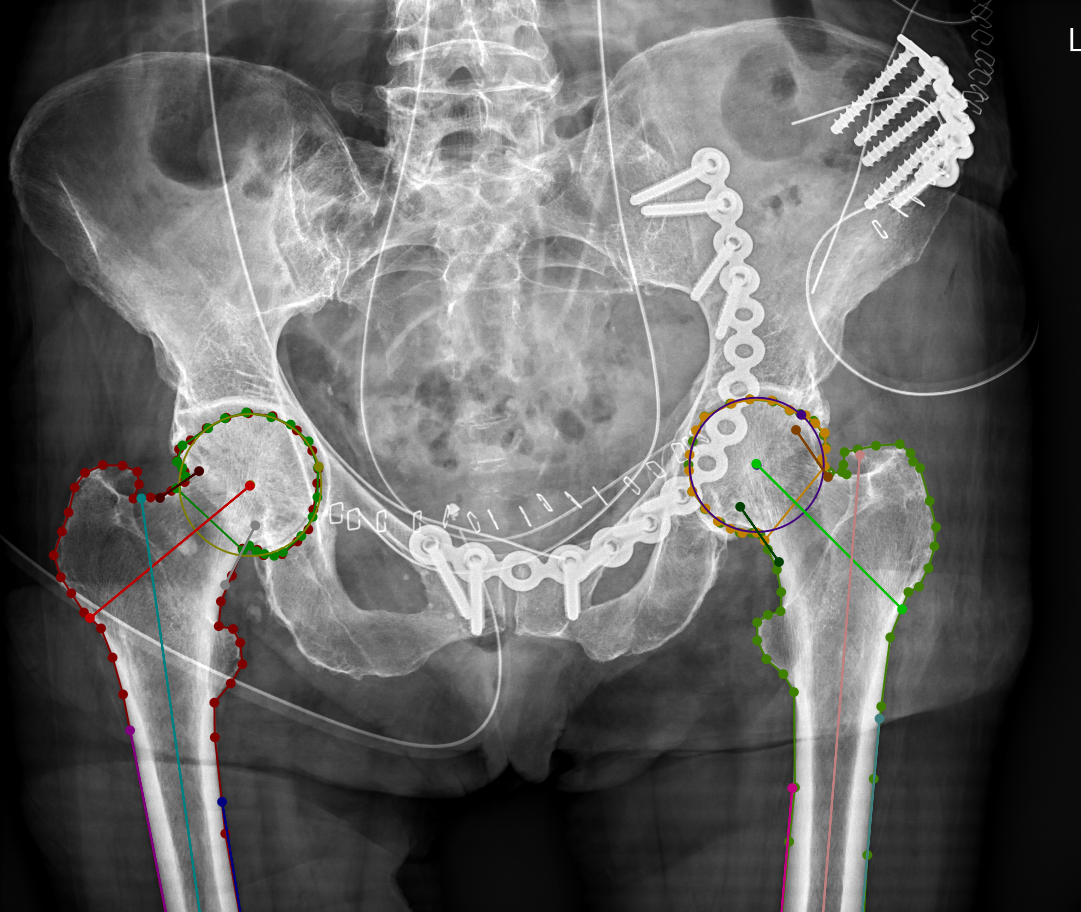}
\caption{Labeled femur} 
\label{fig:annonation}
\end{figure}

\subsection{Training and Implementation}
For the training of the UNet, we created 8:1:1 train, validation, and test split. Since, in few cases, a single patient has more than one X-Rays images, we keep all images of the same patient into the same split to avoid data leak into validation and test split. We also applied colorjitter(contrast) and affine transformations(translate, scale, rotate) in an online augmentations fashion to counter over-fitting. Input was resized (512x512) and normalised before feeding to the network. We used z-score normalisation because most of the pixels were either zero and near to zero except the pixels in the heatmap with greater probabilities. It helped to use sigmoid on predictions to remove the low probability pixels. We used a sigmoid cutoff of \textbf{0.9}.
Since, input mask contains pixel probabilities, we used MSELoss to calculate pixel wise mean squared error.
\newline
Since our input masks were line heatmaps with varying probabilities, We could use pixel-wise crossentropy loss or MSELoss, we selected MSELoss for regression between input mask and predicted mask because segmentation values are continuous and the magnitude of the errors is important to consider.
For the optimization, we used Adam optimizer without weight decay and with learning rate \textbf{8e-5}. We kept the learning rate same for the training as network was able to converge in very few epochs, \textbf{25} in our case. We used the batch size of \textbf{4} as UNet can also generalize well even on smaller batch sizes.

\subsection{Post-processing of Heatmaps}
After predicting the line heatmap with UNet. we apply a sigmoid cut-off to remove points with low probabilities. Then we fit the remaining points via linear regression. Since, outliers can impact the performance of the linear fit, robust linear regression requires outliers suppression. HuborRegressor, TheilSenRegressor and RansacRegressor can be used to fit a robust linear model while suppressing the impact of outliers. RANSAC\cite{RANSAC} performs better when data is highly contaminated by outliers. Therefore, we decided to use RANSAC for outliers removal. 
\newline
RANSAC separates the data points into outliers and inliers groups. Only inliers are used to fit the linear regression model. HuberRegressor can be applied to the inliers set which can further lower the effect of outliers points. 

\begin{figure*}[ht]
\centering
    \includegraphics[width=\linewidth]{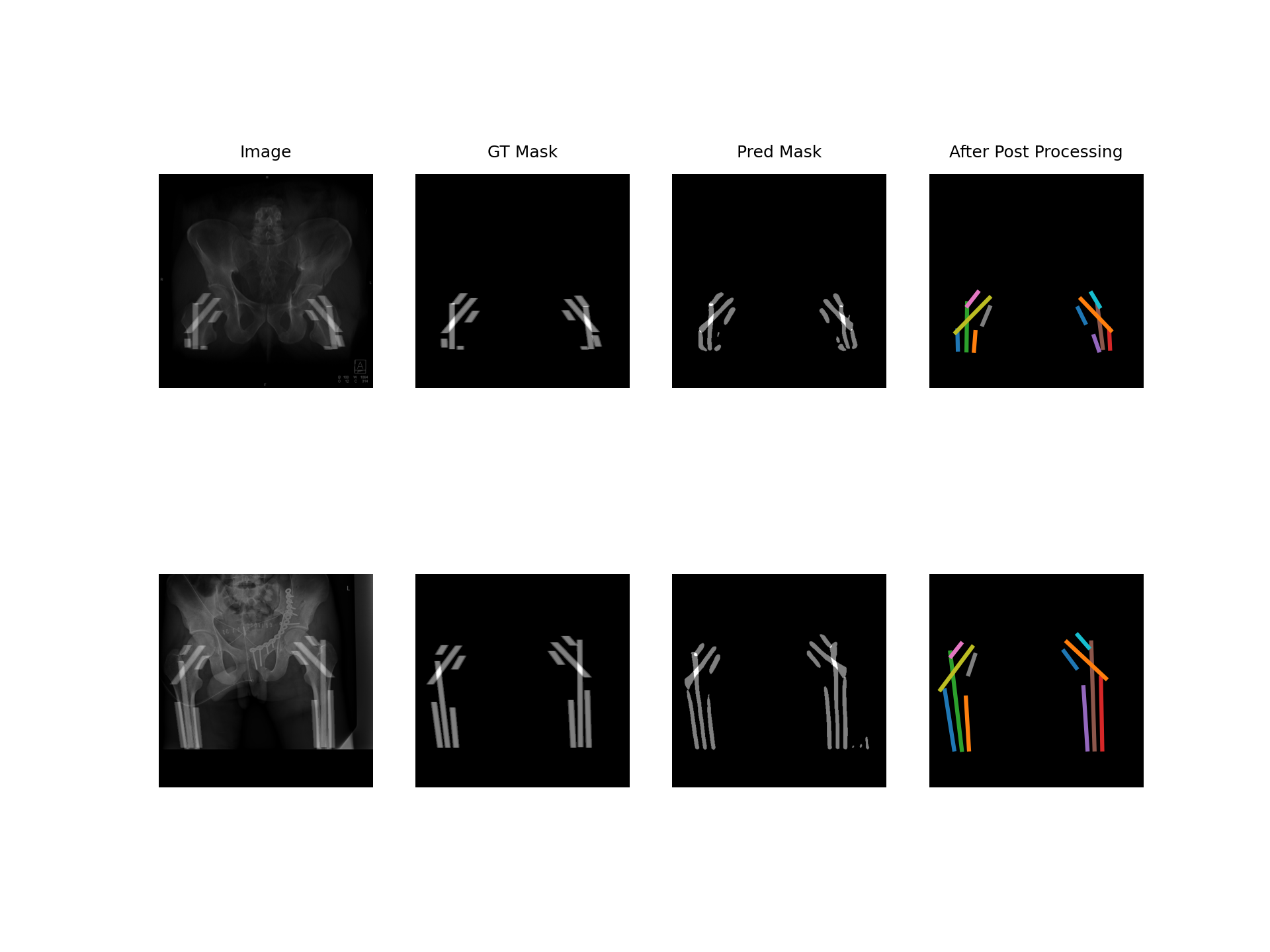}
\caption{1st column contains input X-Rays images with masks overlayed. 2nd column contains input mask that is fed to the UNet. 3rd Column contains masks predicted after the sigmoid cutoff applied at the output. 4th colums contains predicted lines after fitting the linear fit.} 
\label{fig:prediction}
\end{figure*}

\subsection{Evaluation}
We use MSELoss for training the UNet but it cannot be used for model assessment as two models with same loss can have very different accuracy when CCD angle are computed because angle computation includes outliers removal and a linear fit on the inliers using RANSAC.
For the model assessment, we computed euclidean distance between centroids of input mask and predicted mask for individual lines. Only centroid difference is not enough. We also computed the angular error for individual lines to calculate the orientation error. 
\newline
As a final criterion for evaluation, we also computed the final CCD angle for both left and right femurs on test dataset.

\subsection{User Interface}

\subsubsection{Design and Functionality}
\begin{figure*}[ht]
\centering
    \includegraphics[width=0.8\linewidth]{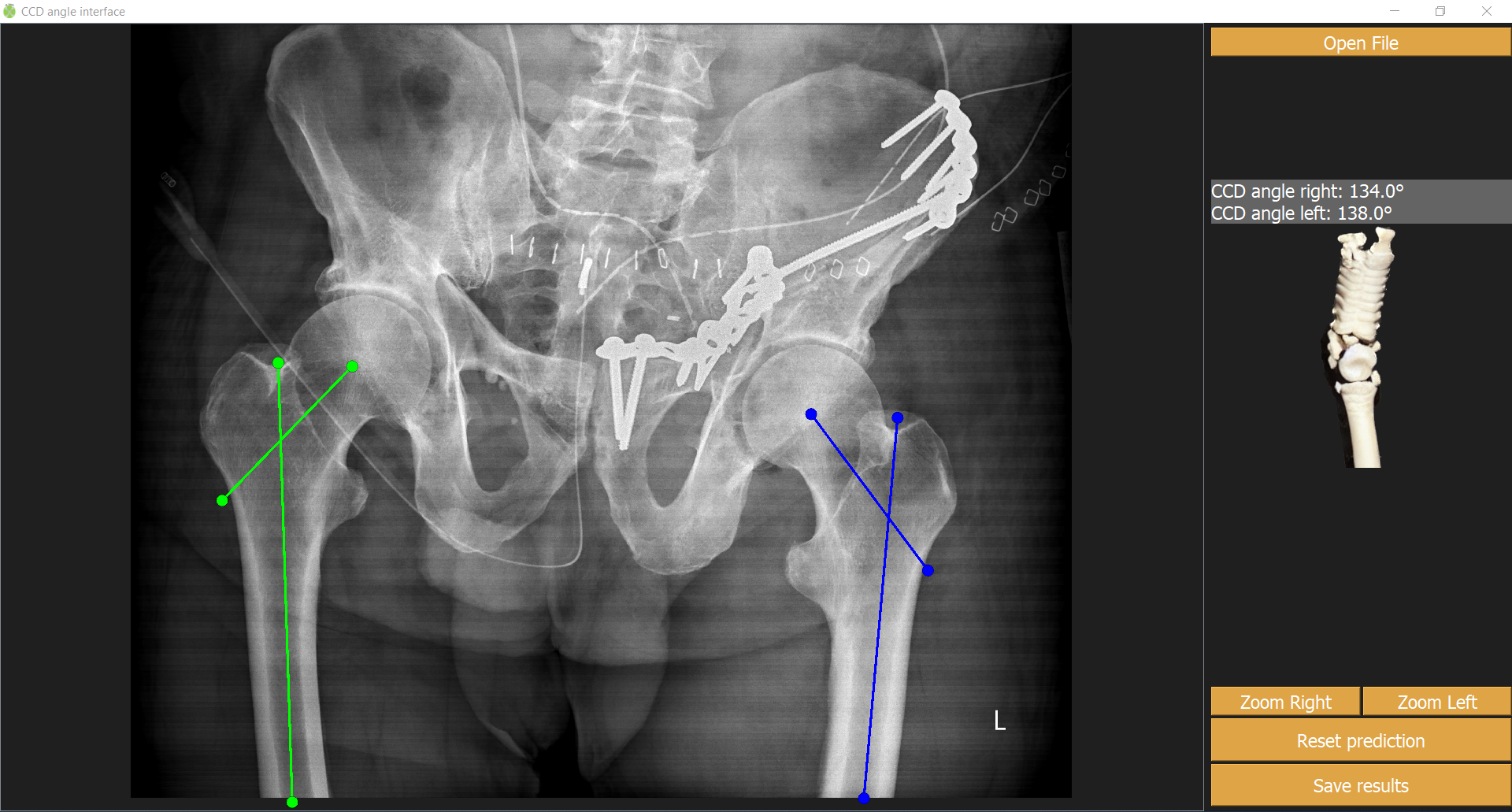}
\caption{Screenshot of user interface with a diagnostic CT scan of the hips opened} 
\label{fig:ui}
\end{figure*}

The interface, which is shown in \cref{fig:ui}, can be roughly divided into two sections. The image and a toolbar.
The central element of the application is the displayed CT image. This is displayed large and takes up 4/5 of the entire screen so that the user can see as much detail of the image as possible. On the right, there is the toolbar, which has all the buttons for the functions and a text field with the CCD angle. The icon that indicates the system state of the voice control function is also shown here, more about this in the next section.

For a successful procedure, it is important to know the angles of both sides of the femur in order to be able to compare them. That´s why the CCD angles are displayed one below the other, to make comparing the numbers as easy as possible. The displayed angle in top is for the right side, because when looking at the CT image from left to right (read-direction), the right hip comes first.
Another option for displaying the angles would have been underneath the image, each on the according side (left/right), but this would separate the two numbers spatially, and capturing both at one glance would not be possible. Therefore we did not opt for this variant.\\
The functionality of the interface is elementary and limited to the essential and most important tasks that are necessary during the intervention. These are intentionally kept simple in order not to overwhelm the user with a flood of new functions and to make the use of the interface as intuitive as possible. This also keeps the necessary training and familiarisation to the tool limited. 
The functions include opening new images, and zooming in on the left or right femur by clicking on a button. After opening, the lines are immediately drawn onto the image, the angle is calculated and displayed in the text box on the right. The lines can be edited by dragging and dropping the endpoints. To make it easier for the user and to draw attention to this option, the colour of the points changes when the mouse is hovered over them. The angle is automatically recalculated with each movement of the lines. 
In addition, a screenshot of the scene can be saved with the image, the lines, the angle and the possibility to add additional text about the case.\\
As the goal of this application is to use it in an intra-operative setting, not only diagnostic images need to be displayed but also intra-operative. Those can be taken for example with a mobile c-arm which has a lot smaller field-of-view, compared to conventional diagnostic CT scanners, which do not cover the entire pelvis and therefore only one side of the hip is displayed. In order to still show both sides, there is the possibility to open two images next to each other. In this case, the picture with the right femur is automatically arranged on the left and the left femur on the right. All functions behave the same as when only one image is open. The only exception is zooming, where the respective picture is maximised.

\subsubsection{Voice control}
A great advantage of our application is that it can be operated without any touch, only by voice. A microphone records what is spoken and executes functions triggered by certain words. All commands that the system recognises are listed in the \cref{tab:voice_commands}. The keywords that are spoken are all intentionally short and intuitive for easy use without having to learn many words.\\
Since a lot is said during an intervention, the system does not always execute the recognised functions, but only if it has been activated beforehand. Activation also happens when a keyword is spoken. When choosing a keyword, it is most important that it is a word that is not used by chance in the operating room, which would accidentally activate the application. Furthermore, the word should be easy to pronounce and simple to remember. Considering these aspects, we decided on the word \textbf{activate}.\\
\begin{figure}[ht]
\centering
    \includegraphics[width=\linewidth]{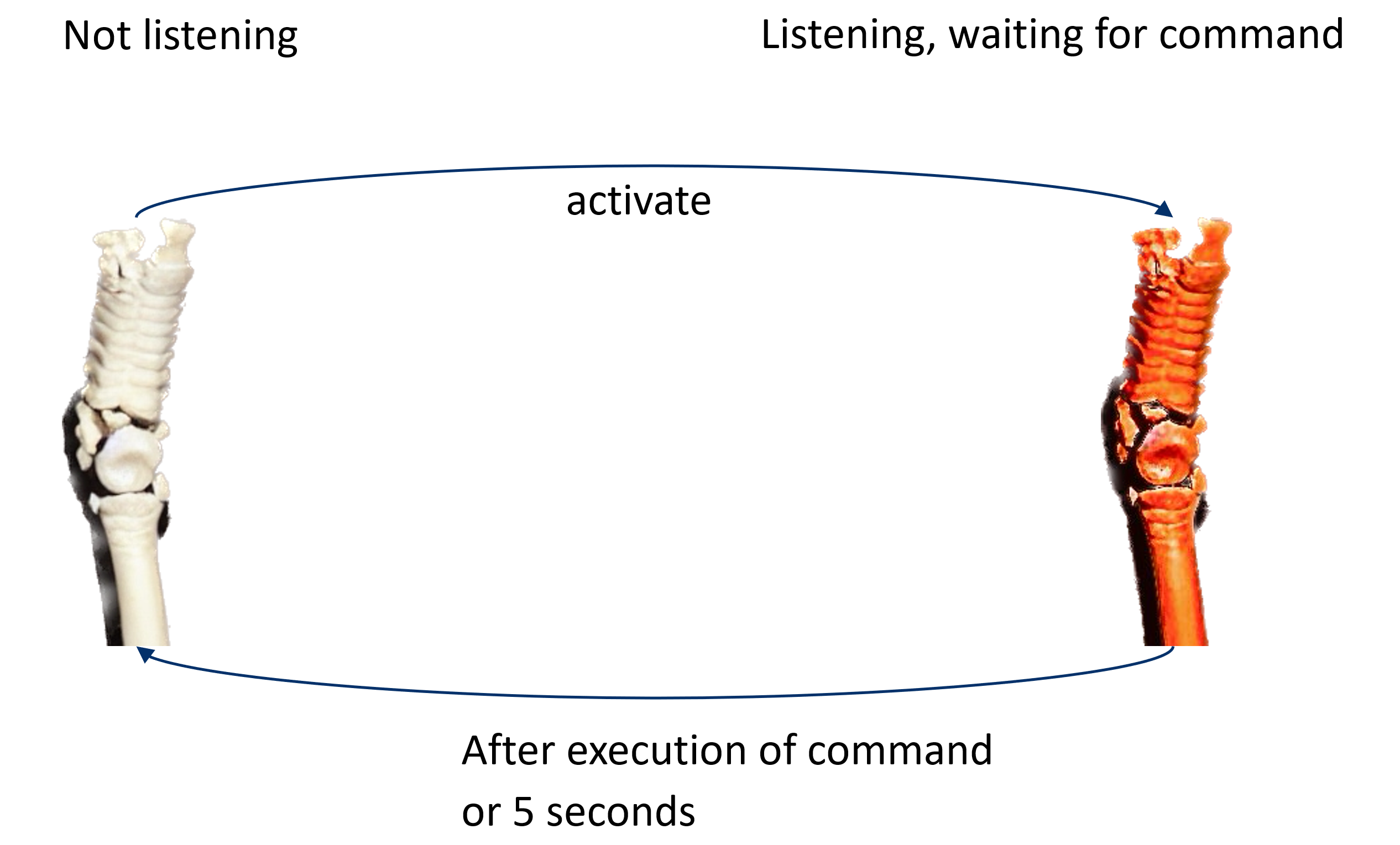}
\caption{System state: The system only executes commands after activation. Once activated the system using the key word \textbf{activate}, the system state will change to be listening and waits until it recognizes any commands. After execution of a command or after waiting for more than 5 seconds it automatically goes back to the default, not listening state.} 
\label{fig:system_state}
\end{figure}
The transition between the states of the system, from "sleeping" to "waiting", is shown in \cref{fig:system_state}. In the default state, the system is "sleeping" and does not react to commands. After activation by saying \textbf{activate}, the system is in a listening state, waiting for further commands, which are then executed. After executing them, the system falls back into the "sleeping" state. Likewise, if none of the cues are recognized within 5 seconds, it goes back into the "sleeping" state. This is because the commands are all very short. Thus, it can be assumed that if no command was recognised within these 5 seconds, the system was either activated by mistake or the command was not recognised correctly. In order to avoid triggering unwanted behaviour, the system falls back into the "sleeping" state.\\
In order for the user to know at all times what state the system is in, an icon was chosen and placed in the center of the toolbar on the right-hand side. If the icon is light grey, the interface is "asleep" and changes its color to bright red as soon as it is activated. This is also shown in \cref{fig:system_state}.\\
When used exclusively with the voice, however, the functionality of the system is slightly limited, as the points of the lines cannot be edited. Otherwise, all functions are possible, just like when using a mouse.

\begin{table*}[ht]
    \centering
    \begin{tabular}{cc}
    \toprule
         command & execution\\
    \midrule
         \textbf{left} & Zoom in on left femur. If two images are open the left image is maximized.\\
         \textbf{right} & Zoom in on right femur. If two images are open the right image is maximized.\\
         \textbf{out} or \textbf{both} & Zoom out to show both sides of the hip at the same time.\\
         \textbf{open} & \multicolumn{1}{p{12cm}}{\centering The The next image is opened from a previously selected folder (before starting the application) sorted by time. The first time the command is used, the latest image is opened.}\\
         \textbf{save} & \multicolumn{1}{p{12cm}}{\centering Waits for additional input (like patient information). When finishing by saying \textbf{ok} function is finished, and a screenshot of the scene with date, time and the dictated patient information is saved to a previously selected (before starting the application) save folder.}\\
    \bottomrule
    \end{tabular}
    \caption{Keywords which are recognized by the speech model and what functionality is triggered by them }
    \label{tab:voice_commands}
\end{table*}

\subsubsection{Implementation}
The user interface was implemented using the python package PyQt which is based on the GUI toolkit Qt. The CT image and the predicted lines are visualized in a QGraphicsScene which allows the user to interact with the lines and update them accordingly. Each line is added to the scene as an item. Each of the lines is the parent item to two endpoints (inherited from QGraphicsPointItem) which are therefore also visible on the scene.
The lines themselves can not be selected and moved, but the two points can. Whenever the lines are updated (via the editing the points), the CCD angle is recalculated, using the formula and the displayed number is updated. \\
Several threads are needed to allow everything to run parallel. The main interface and all updates on the interface are executed in the main thread. When starting the application there are 2 more threads started, which are required to allow speech control.
Communication of the sub-threads with the main thread is done with signals and slots from PyQt.

\subsubsection{User Study Design}
To get a feeling for the usability and feedback on our system, we conducted a user study with our prototype. Four participants were briefly introduced to the medical background and the use of the application and then left alone to test all functions and compare the functionality of using it with mouse and solely with speech recognition.\\
 Since we only have very few participants in our study, all participants used the interface with both voice and mouse. They all used the mouse first to become familiar with the system. 
The study was conducted in a public space, with background noise, to mimic the possible background noise in the operating room.\\
After using the system, all subjects were given a questionnaire. These questionnaires were created using different standardised questionnaires, such as SUS \cite{SUS} and UEQ \cite{UEQ}, and adapted to suite this application. In the first part, statements have to be rated with numbers from 1 (strongly agree) to 5 (strongly disagree). Those questions are on purpose designed differently, some are positives, others have a rather negative tone.
The first part of the questionnaire was answered regarding the interface and functionality in general and the usage with mouse.
In the second part, open questions are asked to give the subjects the possibility of detailed feedback, and to compare the option of using the interface by voice to using the mouse.

\section{Results} \label{Results}
\subsection{Algorithm}
Although we trained our UNet on 12 femur lines, 6 for each side. We ended up just using the shaft and neck centerlines for measuring the CCD angle.
Table III contains performance of neck and shaft centerlines both for left and right femur. Smaller angular error for individual lines indicates the success of out approach in predicting lines. Euclidean distance between centroids is bit larger than expected but it does not contribute much towards final angle calculation as it is calculated before RANSAC.
\newline
Finally, In table II, we compute the CCD angle on test dataset. We got the mean absolute error of 4.3 on left femur and 4.9 on right femur which indicates the success of our approach.
\begin{table}[ht]
    \centering
    \begin{tabular}{cc}
    \toprule
         Femur & Mean Absolute Error (degrees)\\
    \midrule
         Left Femur & 4.3 \\
         Right Femur & 4.9 \\

    \bottomrule
    \end{tabular}
    \caption{Mean absolute error of CCD}
    \label{tab:ccd_pred}
\end{table}

\begin{table*}[ht]
    \centering
    \begin{tabular}{ccc}
    \toprule
         Femur line & Mean Centroids Euclidean distance & Mean Angular error(degrees)\\
    \midrule
         Left shaft centerline & 9.6 & 1.9\\
         Left neck centerline & 14 & 2.7 \\
         Right neck centerline & 7 & 5.0\\
         Right shaft centerline & 12.6 & 2.0\\

    \bottomrule
    \end{tabular}
    \caption{Evaluation of individual femur centerlines}
    \label{tab:lines_evaluation}
\end{table*}

\subsection{User Study}

To evaluate our user study in the usability and user experience of our application, the SUS score was calculated. \cref{tab:sus_scores} shows the SUS scores of the 4 users who were interviewed. All participants seem to be quite satisfied with the application, and achieve a score of 80-90\%. This is clearly above the average of 68\%, which is given in \cite{SUS}. \\
\begin{table}[ht]
    \centering
    \begin{tabular}{ccccc}
    \toprule
          & user 1 & user 2 & user 3 & user 4\\
    \midrule
         SUS score [in \%] & 80 & 85 & 80 & 90\\
         
    \bottomrule
    \end{tabular}
    \caption{SUS score in \% for four user participating in our user study }
    \label{tab:sus_scores}
\end{table}
Even more interesting are the results of the open questions. Many important hints and ideas for further developments were given there. The function of adding and deleting lines if the system does not predict exactly the two centre lines (this happens only very rarely in split screen mode with one femur per side) was introduced. 
The possibility of using the interface with voice was consistently praised by all users, but only one of the users stated that he would prefer to use voice over mouse. The icon that changes color to represent the status of the system was positively highlighted. It was also noted that there might be better words for activation, e.g. a fantasy name. \\
A big concern that was noticed during the user study is that the speech model is biased heavily. The pronunciation is very important, and with a strong accent, the speech recognition feature was almost unusable, as it took many attempts until the correct command was recognized, or in some cases was not recognized at all.

\section{Conclusion}\label{Conclusion}
We proposed a deep learning-based technique for estimating the CCD angle on X-ray images in patients with hip fractures in this report. Our experimental results showed that the proposed method achieved good accuracy results, with an MAE of 4.3 degrees on left femur and 4.9 degrees on right femur. The proposed method has the potential to enhance patient outcomes by assisting in the faster, more accurate and efficient identification of hip disorders.
\newline
One of the main advantages of the proposed method is that it can automatically learn the important features from hip X-ray images to determine the appropriate CCD angle, reducing the need for manual effort. The angle is determined by small changes in bone structure, which can be difficult to discover precisely using standard manual methods, which are also subject to human error and inter-reader variability.
\newline
Another advantage of the proposed approach is that it is simple to integrate into the clinical workflow with the voice command feature in our user interface providing ease of use in critical conditions of an interventional setting for hip fractures correction procedures, allowing for quick and precise calculation and visualizations of the CCD angle from X-ray images. This automated strategy can minimize radiologists' and surgeons' workloads and increase diagnostic and treatment process efficiency.
\newline
Our proposed method, however, has certain drawbacks. One of the constraints is that training requires a big annotated dataset, which may be difficult to get. The dataset we utilized was too small for the model to generalize well. There were also images containing implants, which might have hampered the model's learning process. Moreover, our technique may not generalize well to populations with different demographics or to cases with severe bone deformities, malalignment or images with implants and fractures.

\section{Future Work}  
In future work, we intend to evaluate the performance of the proposed method on a bigger and more diversified dataset. Additionally, we plan to test the proposed method's generalization ability on different types of cases with severe bone abnormalities or malalignment, or cases with implants, and fractures. Lastly, we intend to explore the proposed method's integration into the clinical workflow and assess its influence on the diagnostic process and patient outcomes.

There are a number of aspects of the user interface that can be improved in future versions.
These include more functions such as adding and deleting additional lines. Furthermore medical specialists were wishing for more freedom when opening two images side by side. It would also make sense, for example, to display a diagnostic image showing both sides of the hip together with an intraoperative image showing only one femur. These considerations raise further questions, such as how to then display the angles (since there are now a total of three to be seen). \\
It would also be very interesting to do a larger user study to gain a more detailed insight into how useful the speech recognition function actually is. This could also include other conditions of everyday clinical life, such as typical background noises or a sterile environment.
% \clearpage

% \bibliographystyle{abbrv}
\bibliography{references}

\begin{thebibliography}{10}

\bibitem{hip_fracture}
Kim~Edward LeBlanc, Herbert~L Muncie~Jr, and Leanne~L LeBlanc.
\newblock Hip fracture: diagnosis, treatment, and secondary prevention.
\newblock {\em American family physician}, 89(12):945--951, 2014.

\bibitem{unet}
Olaf Ronneberger, Philipp Fischer, and Thomas Brox.
\newblock U-net: Convolutional networks for biomedical image segmentation,
  2015.
\newblock cite arxiv:1505.04597Comment: conditionally accepted at MICCAI 2015.

\bibitem{he2018mask}
Kaiming He, Georgia Gkioxari, Piotr Dollár, and Ross Girshick.
\newblock Mask r-cnn, 2018.

\bibitem{badrinarayanan2016segnet}
Vijay Badrinarayanan, Alex Kendall, and Roberto Cipolla.
\newblock Segnet: A deep convolutional encoder-decoder architecture for image
  segmentation, 2016.

\bibitem{chen2018encoderdecoder}
Liang-Chieh Chen, Yukun Zhu, George Papandreou, Florian Schroff, and Hartwig
  Adam.
\newblock Encoder-decoder with atrous separable convolution for semantic image
  segmentation, 2018.

\bibitem{liu2021swin}
Ze~Liu, Yutong Lin, Yue Cao, Han Hu, Yixuan Wei, Zheng Zhang, Stephen Lin, and
  Baining Guo.
\newblock Swin transformer: Hierarchical vision transformer using shifted
  windows, 2021.

\bibitem{long2015fully}
Jonathan Long, Evan Shelhamer, and Trevor Darrell.
\newblock Fully convolutional networks for semantic segmentation, 2015.

\bibitem{xiao2018simple}
Bin Xiao, Haiping Wu, and Yichen Wei.
\newblock Simple baselines for human pose estimation and tracking, 2018.

\bibitem{mewes_touchless_2017}
André Mewes, Bennet Hensen, Frank Wacker, and Christian Hansen.
\newblock Touchless interaction with software in interventional radiology and
  surgery: a systematic literature review.
\newblock {\em International journal of computer assisted radiology and
  surgery}, 12(2):291--305, February 2017.

\bibitem{li2020openai}
Rui Li, Kaisheng Zhang, Shuang Li, Minh-Thang Luong, and Quoc~V Le.
\newblock Openai whisper: Speech-to-text transformer.
\newblock {\em arXiv preprint arXiv:2006.04558}, 2020.

\bibitem{RANSAC}
Martin~A. Fischler and Robert~C. Bolles.
\newblock Random sample consensus: A paradigm for model fitting with
  applications to image analysis and automated cartography.
\newblock {\em Commun. ACM}, 24(6):381–395, jun 1981.

\bibitem{SUS}
John Brooke.
\newblock Sus: A quick and dirty usability scale.
\newblock {\em Usability Eval. Ind.}, 189, 11 1995.

\bibitem{UEQ}
Bettina Laugwitz, Theo Held, and Martin Schrepp.
\newblock Construction and evaluation of a user experience questionnaire.
\newblock volume 5298, pages 63--76, 11 2008.

\end{thebibliography}

\end{document}